\documentclass[fleqn,usenatbib]{mnras}
\usepackage{newtxtext,newtxmath,times,verbatim}
\usepackage[T1]{fontenc}
\DeclareRobustCommand{\VAN}[3]{#2}
\let\VANthebibliography\thebibliography
\def\thebibliography{\DeclareRobustCommand{\VAN}[3]{##3}\VANthebibliography}
\usepackage{graphicx}
\usepackage{amsmath}
\usepackage{ulem}
\usepackage{comment}
\usepackage{booktabs}

\title[Variability-selected AGN in massive galaxies]{AGN in massive galaxies identified via optical broadband variability: lessons from VST-COSMOS for future LSST science}

\author[B.Bichang'a]{
B. Bichang'a,$^{1}$\thanks{b.o.bichanga@herts.ac.uk} D. De Cicco,$^{2,3,4}$ 
S. Kaviraj$^{1}$,
I. Lazar$^{1}$,
A. Watkins$^{1}$,
G. Martin$^{5}$ and
D. Kakkad$^{1}$
\\
$^{1}$Centre for Astrophysics Research, Dept of Physics, Astronomy and Mathematics, University of Hertfordshire, Hatfield, AL10 9AB, UK\\
$^{2}$Department of Physics, University of Napoli “Federico II”, Via Cinthia 9, 80126 Napoli, Italy\\
$^{3}$Millennium Institute of Astrophysics (MAS), Nuncio Monse\~nor Sotero Sanz 100, Providencia, Santiago, Chile\\
$^{4}$INAF - Osservatorio Astronomico di Capodimonte, via Moiariello 16, 80131 Napoli, Italy\\
$^{5}$School of Physics and Astronomy, University of Nottingham, University Park, Nottingham NG7 2RD, UK\\
}

\pubyear{\the\year{}}

\begin{document}
\label{firstpage}
\pagerange{\pageref{firstpage}--\pageref{lastpage}}
\maketitle

\begin{abstract}
We study the properties of 56 massive (M$_{\rm{\star}}$ > 10$^{10}$ M$_{\odot}$) galaxies at $z<1$ that host AGN, detected via their broadband optical variability in the VST-COSMOS survey. VST-COSMOS provides a nearly-identical single visit depth ($r$ $\sim$ 24.6 mag) and temporal baseline (eleven years) as the forthcoming Legacy Survey of Space and Time (LSST), albeit in a much smaller 1 deg$^2$ footprint (four orders of magnitude smaller than that of the LSST). We compare the properties (morphologies, the presence of interactions, rest-frame colours and environment) of our AGN to galaxies in a control sample, which are drawn from the non-variable population and matched in redshift and stellar mass to their AGN counterparts. The fraction of AGN with early-type morphology ($\sim$55 per cent) and the fraction that is interacting ($\sim$23 per cent) are similar to what is observed in the controls, suggesting that these AGN are not primarily triggered by interactions. Similarly, the AGN and controls do not show strong differences in their rest-frame $(u-z)$ colours or local environment, suggesting that neither the recent star formation histories nor the surroundings of the AGN are strongly atypical of the general galaxy population. This study provides a glimpse into forthcoming AGN science using the LSST. With vastly improved statistics, LSST will offer unprecedented insights into AGN demographics, host-galaxy evolution and the processes that fuel supermassive black holes, potentially reshaping our understanding of their place in the Universe.
\end{abstract}



\section{Introduction}
\label{sec:intro}

Supermassive black holes (BHs) are thought to be ubiquitous in massive galaxies. These BHs accrete material from their surroundings, turning into active galactic nuclei (AGN) that release vast amounts of energy back into the interstellar medium via radiation-pressure driven outflows and/or collimated radio jets (see \citet{Fabian_2012} for a review). The broad impact of this process is to either heat or remove the gas from a galaxy's potential well which reduces the level of star formation activity \citep[e.g.][]{Cattaneo2009,Heckman2014,Beckmann2017}. AGN can also impact the wider environment by heating the gas surrounding galaxies and preventing further accretion \citep{Dubois2013}. In many cosmological simulations \citep[e.g.][]{Hopkins2014,Schaye2015,Kaviraj2017,Pillepich2018,Dave2019,Dubois2021}, this negative AGN feedback is invoked to bring mock galaxy populations into agreement with empirical trends, such as the observed colour bimodality of galaxies \citep[e.g.][]{Strateva2001,Kauffmann2003,Baldry2004, Wyder2007, Whitaker2011} in which red and blue galaxies broadly map on to quenched and star forming systems respectively \citep[e.g.][]{Brinchmann2004,Faber2007,Bluck2014}. Furthermore, both the theoretical and observational literature suggests a tight correlation between BH mass and properties of the host galaxy, such as stellar mass \citep[e.g.][]{Marleau2013,Martin2018} and stellar velocity dispersion \citep[e.g.][]{Haring2004,Kormendy2013} implying a close link between the growth of the BH and that of its host galaxy. 

AGN can be selected using various methods across the electromagnetic spectrum, with different selection techniques finding varying samples of AGN that often do not overlap \citep[e.g.][]{Hickox2018,Lyu2022}. For example, a common selection technique in the optical wavelengths uses the ratio of forbidden and allowed emission lines to identify systems where the gas ionization may be dominated by the AGN rather than star formation \citep[e.g.][]{Baldwin1981,Kewley2001,Mezcua2024}. X-ray \citep{Ranalli2005} and near-infrared (NIR) to mid-infrared (MIR) colour selection methods \citep[e.g.][]{Donley2012} can be adopted to obtain samples of obscured AGN \citep[e.g.][]{Padovani2017}, which are largely missed in optical/UV selection techniques. While X-ray AGN samples include unobscured AGN, they are biased against X-ray faint objects, tend to miss a large fraction of Compton-thick AGN \citep[e.g.][]{Donley2012, Buchner2021} and result in more massive galaxies than those selected in the infrared wavelengths \citep[e.g.][]{Zhiyuan2022}. The NIR -- MIR selection methods may also struggle to separate light from the AGN from that of the host galaxy, especially in less powerful AGN \citep[e.g.][]{Donley2012}. 

Variability selection is driven by the fact that the accretion on to the BH is inherently time variable. This, in turn, induces variability in the AGN flux across a range of wavelengths \citep[e.g.][]{Ulrich1997,Elmer2020} and serves as a useful addition to the suite of existing AGN selection methods which can select both unobscured and obscured AGN (although the sensitivity to obscured AGN can be lower, at least in the optical wavelengths). For example, \citet{Decicco2019} have used optical broadband variability to recover 256 out of 299 known AGN in bright (i.e. massive) galaxies, achieving 82 and 18 per cent completeness rates for unobscured and obscured AGN respectively. Their sample, which is similar to that selected by \citet{Kimura2020}, shows significant overlap with X-ray and infrared-selected AGN, indicating that optical variability is complementary to other AGN selection methods even when higher flux thresholds are used.  

The variability of AGN is observed to be stochastic and aperiodic, which is also evident from their power spectra which have no distinct peaks or features \citep{Kelly2009}. The amplitude of the variability is inversely correlated with nuclear luminosity \citep[e.g.][]{Hook1994, Vanden2004, Ponti2012, Zuo2012, Kozlowski2016, Simm2016} and rest-frame wavelength, with larger amplitudes observed at shorter wavelengths \citep{Cutri1985, helfand2001, Paltani&Courvoisier1994, Vanden2004, Trevese2008, Macleod2010, Zou2012,Li2018, Morganson2014, sanchez2018}. The variability at shorter wavelengths, e.g. in the X-rays, takes place on shorter timescales \citep[]{Vaughan2011}, compared to that in the UV/optical regime \citep{Padovani2017, sanchez2018}. The shorter timescales observed in the X-rays are consistent with the emission originating from the corona \citep{Haardt1991, Lawrence1987, Ponti2012}, while the optical emission originates from the optically thick accretion disc \citep{scott2024}. The underlying physical mechanisms that are responsible for the observed variability are, however, not well understood. 

Samples of optical variability-selected AGN tend to comprise moderate to low-luminosity AGN \citep[$L_{\rm bol}$ $< 10^{42}$ $\text{erg s}^{-1}$, e.g.][]{Saikia2018} that are largely unobscured \citep[e.g.][]{Bershady_1998, Villforth2012, Pouliasis2019}. While relatively short temporal baselines of a few years can identify optical variability-selected AGN, longer baselines (e.g. of around a decade) are typically required to obtain more complete samples \citep[e.g.][]{Decicco2015, Decicco2019}. The need for sufficiently long temporal baselines is further supported by \citet{Hu2024}, who investigate the dependence of optical variability-selected AGN samples on the temporal baselines employed, assuming AGN light curves are characterized by the damped random walk model \citep{Kelly2009}. They report that sufficient temporal baselines of at least 10 times the damping timescale (i.e. the time required for the AGN light curve to become uncorrelated) is important to retrieve unbiased variation timescales. \citet{De-Cicco2022}, who use the structure function to analyse AGN light curves, also find that sufficiently long baselines are necessary to identify larger fractions of obscured AGN. 

The advent of the Legacy Survey of Space and Time \citep[LSST; ][]{Ivezic2019} in the near future offers a route to identifying millions of AGN via optical broadband variability. LSST will provide a three-day cadence, a single-visit depth of $r\sim24.7$ mag, an 18,000 deg$^2$ footprint and a ten-year temporal baseline when it is complete. The detailed census of variable AGN that will be possible to create using LSST will greatly improve our understanding of accretion mechanisms of supermassive BHs and provide insights into BH demographics by enabling us to study low-luminosity AGN and those that originate from low-mass BHs in dwarf galaxies \citep[e.g.][]{Kimura2020}. 

In this paper, we use optical broadband variability to identify AGN in massive (M$_{\rm{\star}}$ > 10$^{10}$ M$_{\odot}$) galaxies out to intermediate redshift ($z\sim1$), using the VST-COSMOS survey, which represents a unique precursor to LSST. VST-COSMOS offers a virtually identical single-visit depth and temporal baseline to LSST, albeit within a 1 deg$^2$ area. While our results are interesting in their own right, crucially they provide a preview of how LSST can be used to perform AGN science when it begins providing data.

The plan for this paper is as follows. In Section 2 we describe the construction of a sample of massive galaxies that host variability-selected AGN, the morphological classification of these objects using visual inspection and the estimation of distances to the nearest filaments within the cosmic web, which we use as our measure of local environment. We also describe the construction of control samples matched in stellar mass and redshift that we use to compare the properties of our AGN to that of the general massive-galaxy population. In Section 3 we first discuss the completeness of our AGN sample and then compare the properties (morphologies, the presence of interactions, rest-frame colours and environment) of our AGN to that in the general galaxy population. We summarise our findings in Section 4.


\section{Data}

\subsection{Identifying variable sources in the VST-COSMOS survey}
\label{sec:variable_sources}

The sample of variable galaxies that forms the basis of our study is constructed using observations from the VST-COSMOS survey which was performed using OmegaCAM \citep{Kuijken2011} on the VST \citep{Capaccioli2011} at the Cerro Paranal observatory. The variable catalogue construction follows \citet{Decicco2015} and \citet{Decicco2019} and is briefly described here. 

Variable sources are extracted from 68 $r$-band observations of the COSMOS field over a period of 11 years. A threshold for selecting variable objects is set by computing the average $r$-band magnitude over the 11-year baseline and the corresponding root mean square (RMS) deviation of each source from its lightcurve. Sources are then defined to be variable if their RMS deviations are in excess of the 95th percentile of the distribution of these values. To identify spurious objects, each variable source candidate is visually inspected and given a quality label from 1 to 3, where a flag of 1 indicates a source without any issues, a flag of 2 indicates a source that is highly likely to be variable but has a neighbour and a flag of 3 corresponds to cases which have a close neighbour which could affect the flux of the source in question (making its variability potentially spurious). Although the single-visit depth of VST-COSMOS is 24.6 mag, we limit our analysis to sources brighter than 23.5 mag, to further minimize any contamination from noisy objects (faint objects that could be affected by background noise or have low signal-to-noise which would result in detection issues). Finally, following \citet{Decicco2015} and \citet{Decicco2019}, we restrict our study only to sources which have a flag of 1 or 2.

AGN selection via optical variability not only identifies typical AGN but is also able to identify low-luminosity AGN \citep[e.g.][]{Villforth2012}, which tend to have a higher variability amplitude, due to the anti-correlation between variability and luminosity \citep[e.g.][]{deVries2005,Gallastegui-Aizpun2014} and which can be missed when using other selection methods such as colour and X-ray to optical flux ratio selection \citep[e.g.][]{Boutsia2009,Choi2014}. Such a population is therefore useful for providing a holistic census of AGN and exploring outstanding questions on AGN triggering mechanisms and the quenching of star formation via AGN feedback \citep{Heinis2016}.


\subsection{A catalogue of massive variable galaxies}
\label{variable_catalogue}

The second step in our sample construction is to identify massive galaxies in the COSMOS field \citep{Scoville2007}. We construct this sample using the Classic version of the public COSMOS2020 catalogue \citep{Weaver2022}. COSMOS2020 applies the \textsc{LePhare} SED-fitting algorithm \citep{Arnouts2002,Ilbert2006} to deep UV to mid-infrared photometry in around 40 broad and medium-band filters in order to calculate physical parameters for individual galaxies (e.g. stellar mass, redshift and rest-frame colours). The photometry is derived from the following instruments: GALEX \citep{Zamojski2007}, MegaCam/CFHT \citep{Sawicki2019}, ACS/HST \citep{Leauthaud2007}, Subaru/Hyper Suprime-Cam \citep{Aihara2019}, Subaru/Suprime-Cam \citep{Taniguchi2007,Taniguchi2015}, VIRCAM/VISTA \citep{McCracken2012} and IRAC/Spitzer \citep{Ashby2013,Steinhardt2014,Ashby2015,Ashby2018}. It is worth noting that the COSMOS2020 object detection includes optical ($i,z$) data from the ultra-deep layer of the Hyper Suprime-Cam Subaru Strategic Program (HSC-SSP), which has a point-source depth of $\sim$28 mag \citep{Aihara2019}, $\sim$10 mag deeper than the magnitude limit of the SDSS spectroscopic main galaxy sample \citep[MGS; e.g.][]{Alam2015}. The accuracy of the photometric redshifts in COSMOS2020 exceeds 1 and 4 per cent for bright ($i<22.5$ mag) and faint ($25<i<27$ mag) objects respectively. Here we use the stellar masses, photometric redshifts and rest-frame colours provided by COSMOS2020. 

To construct a sample of massive galaxies we select objects that are classified as galaxies by \textsc{LePhare}  (`type' = 0 in the COSMOS2020 catalogue) and also as `extended' (i.e. galaxies) by the  HSC pipeline in the $g$, $r$, $i$ and $z$ filters. We then select galaxies which (1) have stellar masses in the range M$_{\rm{\star}}$ > 10$^{10}$ M$_{\odot}$, (2) have redshifts in the range $z<1$ (the redshift limit to which we can construct complete, unbiased samples of massive galaxies given the detection limit of VST-COSMOS), (3) lie within the HSC-SSP footprint and outside bright-star masks and (4) have both $u$-band and mid-infrared photometry (since a long wavelength baseline produces more accurate physical parameters, see e.g. \citet{Ilbert2006}). Note that X-ray sources are excluded from our analysis. While we do not impose an initial threshold for the goodness of fit in our galaxy selection -- which is provided as the reduced chi-squared of the best-fit template from the \textsc{LePhare} SED fitting -- we consider the acceptability of the fits following the procedure outlined in \citet{Smith2012}. A fit is considered to be acceptable if its best-fitting $\chi^2$ is below the 99 per cent confidence threshold for the given number of photometric bands. This check indicates that the fit quality for all galaxies used in this study is acceptable and none of the objects falls within the outlier region of the chi-squared distribution. Finally, this massive galaxy sample is cross matched with the sample of variable systems described above, producing 56 AGN that reside in massive galaxies, within the stellar mass and redshift ranges described above. 

\subsection{Control samples}
\label{sec:control_sample}

To compare the properties of our optical variability-selected AGN to the general galaxy population, we construct control samples of non-AGN (defined as galaxies that are classified as non-variable in the procedure described in Section \ref{sec:variable_sources}) which are matched in stellar mass and redshift to their AGN counterparts. The matching is performed within tolerances of 0.05 in redshift and 0.05 dex in stellar mass. Each AGN is assigned five control galaxies and no control object is assigned to more than one AGN. 

For the comparison between AGN and non-AGN we only consider objects in the part of the redshift vs stellar mass parameter space where galaxies are mass-complete at the depth of the VST-COSMOS survey (see Section \ref{sec:completeness}). This is because galaxies outside the completeness region of a survey are likely to have elevated levels of star formation which can introduce biases in terms of colour and morphology \citep{Kaviraj2025}. 

\begin{figure}
    \centering
    \includegraphics[width=0.9\linewidth]{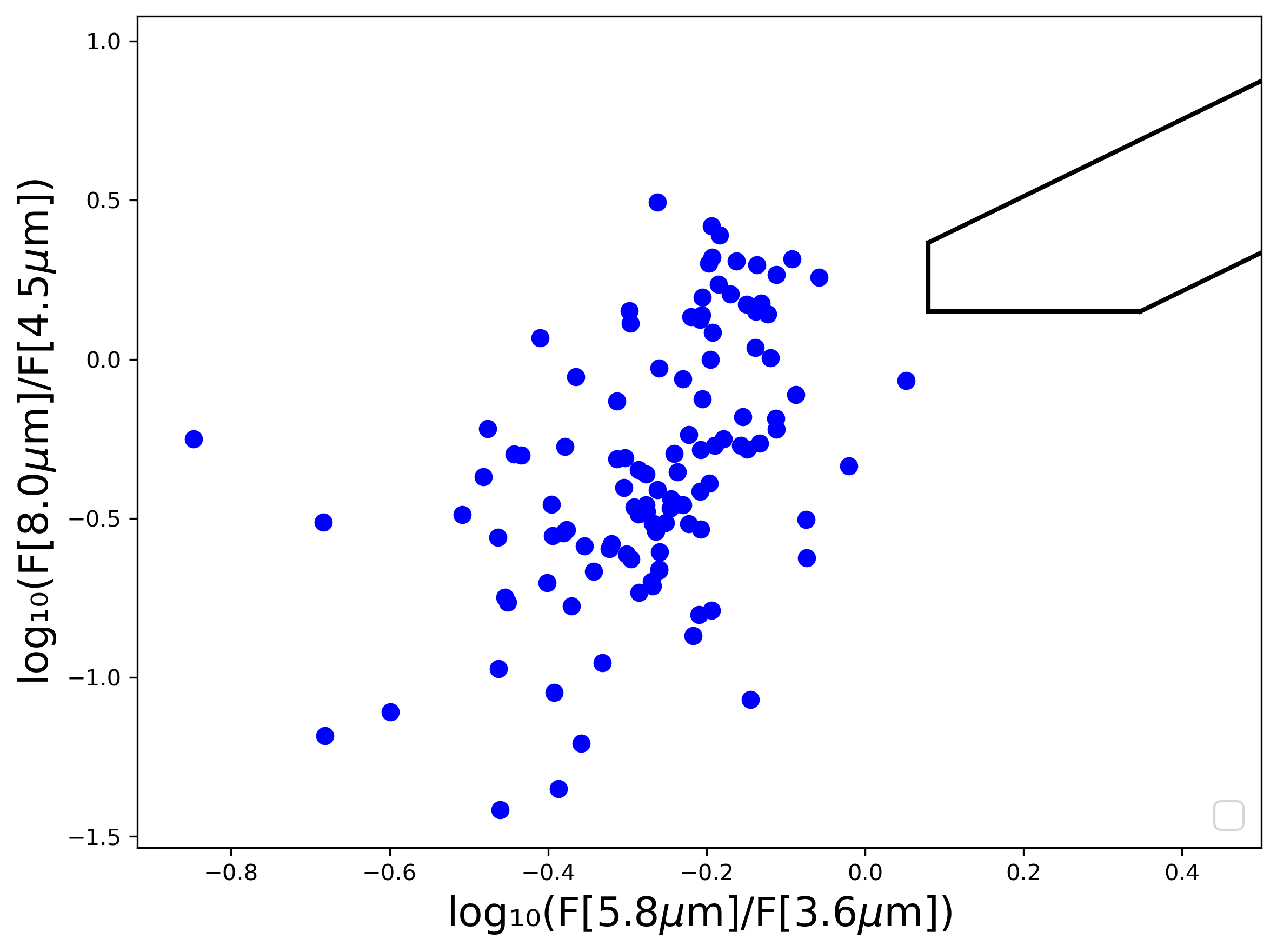}
    \caption{Mid-infrared diagram where colours are obtained as ratios of the fluxes from the four channels of the Spitzer Large Area Survey with Hyper-Suprime-Cam (SPLASH; \citealt{Weaver2022}). The blue dots indicate the sources in our control sample. The solid line defines the AGN locus based on \citet{Donley2012}.
}
    \label{fig:mir}
\end{figure}

Obscured AGN are likely to make up the majority of the total AGN population and we note that AGN selection via variability typically identifies around a fifth of the population of obscured AGN \citep[e.g.][]{Decicco2019}. It is therefore worth considering if our control samples could be contaminated by obscured AGN. In this context, we note first that, as mentioned above, X-ray sources are not considered in this study. Furthermore, we find that our control sample falls outside the mid-infrared AGN colour selection region in \citet{Donley2012}, indicating that the controls are unlikely to be contaminated by large numbers of unobscured AGN (see Figure \ref{fig:mir}).

\subsection{Classification of galaxy morphologies}
\label{sec:morphology_classification}

We classify galaxy morphology via visual inspection of the F814W ($I$-band) images \citep{Koekemoer2007,Massey2010} of our variable and control galaxies from the Advanced Camera for Surveys (ACS) instrument on board the \textit{Hubble Space Telescope (HST)}. A range of techniques has been used in the recent literature to classify galaxy morphologies, ranging from visual inspection to automated parametric \citep[e.g.][]{Conselice2003,Lotz2004} and non-parametric techniques \citep[e.g.][]{Martin2020,Uzeirbegovic2020,Walmsley2020}. However, automated techniques are benchmarked against visual inspection, which is typically considered the most accurate form of morphological classification \citep[e.g.][]{Kaviraj2014b,Lintott2011}. While automated techniques are needed when the size of datasets is large (and visual inspection becomes prohibitively time consuming), our relatively small sample size is well suited to the use of visual inspection for the morphological classification process. 

The visual inspection is performed by one expert classifier (SK). The galaxies are classified into two very broad morphological classes: early-type galaxies (ETGs) and late-type galaxies (LTGs). Following the classical definition of these morphological types, galaxies that exhibit central light concentrations surrounded by smooth ellipsoidal light distributions are classified as ETGs, while galaxies that do not exhibit these characteristics are classified as LTGs. Since we will be interested in exploring the impact of interactions on AGN activity, we also flag systems that show clear evidence of an ongoing or recent interaction. These interaction signatures manifest themselves in the form of tidal features or internal asymmetries, both of which indicate a post-interaction remnant, or tidal bridges with another system which indicate that the galaxy in question is currently interacting with another system. Note that we do not select interacting systems by simply considering galaxies that show proximity in spatial coordinates or redshift. Rather we rely on the presence of the features described above, which firmly signpost the presence of an ongoing or recent interaction \citep[e.g.][]{Kaviraj2014b}. 

\begin{figure*}
\center
\includegraphics[width=2\columnwidth]{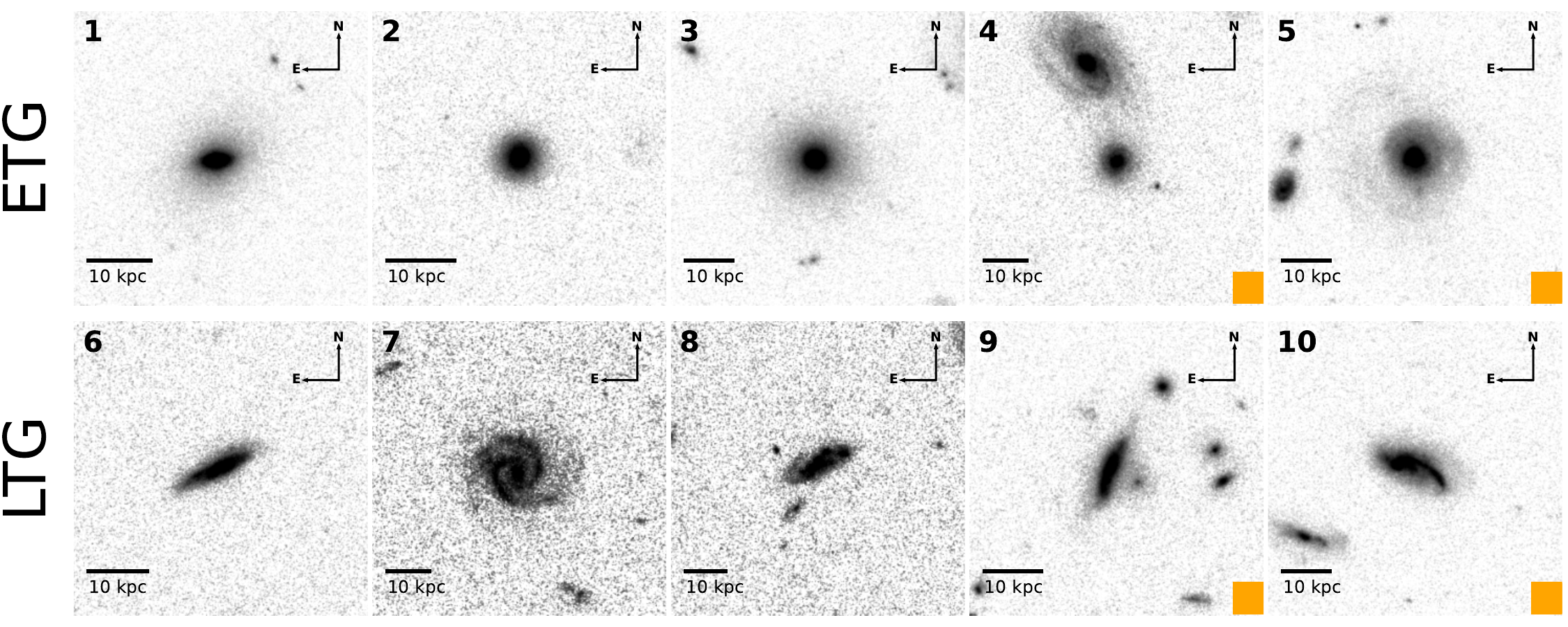}
\caption{Example images of galaxies in our study (each image is 10 x 10 arcsec$^2$). The images are taken using the HST-ACS F814W ($I$-band) filter. The top and bottom rows show early-type galaxies (ETGs) and late-type galaxies (LTGs) respectively. Interacting systems are indicated using an orange filled square in the lower right-hand corner of the image. Galaxy 4 is involved in an ongoing merger with a companion to the north and exhibits a tidal bridge. Galaxy 5 is a post-merger system which shows tidal features to the west. Galaxies 9 and 10 both appear visibly asymmetric due to a recent or ongoing interaction.} 
\label{fig:images}
\end{figure*}

Figure \ref{fig:images} presents example images of our ETGs (top row) and LTGs (bottom row), with interacting systems indicated using an orange filled square in the lower right-hand corner of the image. Galaxy 4 is involved in an ongoing merger with a companion to the north and exhibits a tidal bridge. Galaxy 5 is a post-merger system which shows tidal features to the west. Galaxies 9 and 10 both appear to be accreting a smaller companion in their central regions and are visibly asymmetric. 

\subsection{Local environment}
\label{sec:distance_to_filaments}

To probe the effect of local environment we use DisPerSE \citep{Sousbie2011}, a structure-finding algorithm, to measure the locations of our galaxies in the cosmic web. DisPerSE first uses the positions of individual galaxies to compute the ambient density field via Delaunay tessellations \citep{Schaap2000}. Following \citet{Lazar2023} and \citet{Bichanga2024}, who have performed an identical analysis using DisPerSE, we use the accurate COSMOS2020 redshifts to define relatively narrow redshift slices ($\Delta z \sim 0.03$) in which to build density maps. We only use the positions of massive (M$_{\rm{\star}}$ > 10$^{10}$ M$_{\odot}$) galaxies to create these maps because they have the smallest redshift errors and dominate their local gravitational potential wells. Each galaxy is weighted by the area under its redshift probability density function that is contained within the redshift slice in question. 

DisPerSE uses the density map thus constructed to identify the locations of critical points i.e. minima, maxima and saddles (which correspond to the centres of filaments). The saddle points are then connected to the local maxima, forming a set of ridges that describes the network of filaments in the cosmic web. To remove ridges that may be spurious and close to the noise, a `persistence' parameter is used to set a threshold value for retaining pairs of saddles and maxima within the density map. A persistence of `\textit{N}' causes all saddle-maxima pairs which have Poisson probabilities below \textit{N}$\sigma$ from the mean to be removed. Here, we set the persistence equal to 2, following the methodology of \citet{Laigle2018} and \citet{Bichanga2024}, who have implemented DisPerSE on redshift slices of similar widths as in our analysis. In this study, we consider the projected distance to the nearest filament (i.e. the distance to large-scale structure) as our measure of local environment. We refer readers to \citet{Sousbie2011} for further details about the algorithm.


\section{Properties of massive galaxies that host AGN}
\label{sec:properties_of_AGN}

\subsection{Completeness}
\label{sec:completeness}

\begin{figure}
    \centering
    \includegraphics[width=0.9\linewidth]{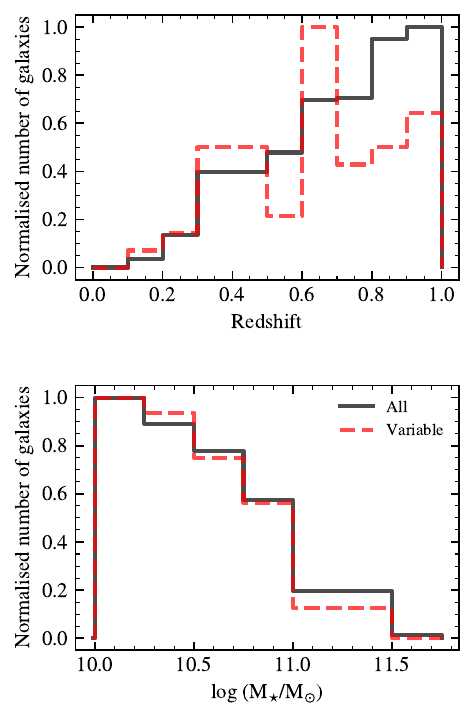}
    \caption{Distributions of redshift (top) and stellar mass (bottom) of our variability-selected AGN (red) and massive galaxies from the parent COSMOS2020 catalogue (black). Both samples have been normalised to a peak value of 1.}
    \label{fig:completeness}
\end{figure}

Figure \ref{fig:completeness} compares the mass and redshift distributions of our AGN to that of all galaxies in the COSMOS2020 catalogue in the same stellar mass and redshift range as the AGN population. While the stellar mass distributions are similar, the detectability of variable sources generally decreases with redshift due to surface brightness dimming and because the relatively low spatial resolution of ground-based surveys like VST-COSMOS generally results in lower signal-to-noise, reducing the chances of detecting nuclear variability. 

\begin{figure}
    \centering
    \includegraphics[width=\linewidth]{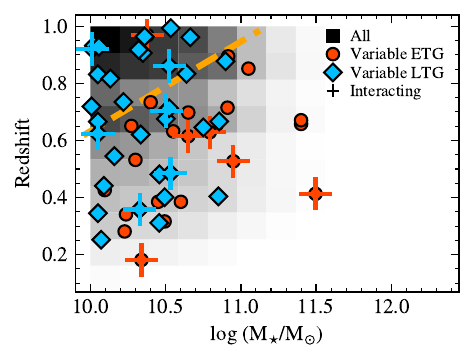}
    \caption{Redshift vs stellar mass for our galaxies. The heatmap indicates the parent COSMOS2020 galaxy population. AGN are shown using filled symbols, with the different morphological classes shown colour-coded (see legend). Interacting galaxies are indicated using crosses. The dashed orange line indicates the redshift at which a galaxy population of a given stellar mass is likely to be mass-complete at the depth of the VST-COSMOS survey (see text in Section \ref{sec:completeness} for details).}
    \label{fig:redshift_mass_completeness}
\end{figure}

Figure \ref{fig:redshift_mass_completeness} presents the redshift vs stellar mass parameter space for our galaxies. The dashed orange line in this figure indicates the redshift at which a galaxy population of a given stellar mass is likely to be mass-complete at the depth of the VST-COSMOS survey. Following \citet{Kaviraj2025}, this is defined as the redshift at which a purely-old `simple stellar population' (SSP) of a given stellar mass, that forms in an instantaneous burst at $z=2$, is detectable, at the depth of VST-COSMOS. This purely-old SSP can be used as a faintest limiting case, since galaxies in the real Universe, which are unlikely to be composed uniquely of old stars, will be more luminous than this value. Thus, if this limiting case is detectable at the depth of a given survey, then the entire galaxy population at a given stellar mass will also be detectable. 

The parameter space below the dashed orange line therefore denotes the region where the VST-COSMOS galaxy population is mass-complete under this conservative assumption. As noted in \citet{Kaviraj2025}, galaxies outside the completeness region of a survey are likely to exhibit elevated levels of star formation and may, therefore, show biases in colour and morphology. Indeed, Figure \ref{fig:redshift_mass_completeness} is consistent with this notion, given that the galaxies above the dashed orange line are overwhelmingly dominated by late-type systems, which typically have higher levels of star formation than their early-type counterparts \citep[e.g.][]{Kaviraj2014}.

\subsection{Morphology and the role of interactions}
\label{sec:morphology_and_interactions}

\begin{table}
\begin{center}
\begin{tabular}{ c | c | c }
\toprule
& AGN & Control\\
\midrule
ETG fraction & 0.55 $\pm$ 0.08 & 0.48 $\pm$ 0.04\\
Interacting fraction & 0.23 $\pm$ 0.06 & 0.16 $\pm$ 0.03\\
\end{tabular}
\end{center}
\caption{Comparison of the fraction of ETGs (first row) and interacting galaxies (second row) between the AGN (second column) and control (third column) populations. The uncertainties are estimated following \citet{Cameron2011}.}
\label{tab:morphology}
\end{table}

Table \ref{tab:morphology} compares the fractions of AGN (second column) and control galaxies (third column) that are early-type (first row) and interacting (second row) respectively. The uncertainties are estimated following \cite{Cameron2011}, who calculate Bayesian binomial confidence intervals using the quantiles of the beta distribution. These are more accurate than simpler techniques, like using the normal approximation, which tends to misrepresent the statistical uncertainty under the sampling conditions typically encountered in astronomical surveys. Within the uncertainties, the two populations do not show strong differences in either the ETG or the interacting fractions. Employing a z-test to evaluate the similarity of the ETG and interacting fractions yields p-values of 0.43 for the ETG fractions and 0.16 for the interacting fractions, respectively. Since both p-values exceed the conventional significance level of 0.05, we cannot reject the null hypothesis that the two distributions are comparable. It is worth noting here that, given that our sample size is relatively small, an analysis using a larger sample (e.g. using the LSST) is needed to put this analysis on a firmer statistical footing.

We note that the interpretation of the AGN interaction fraction requires some caution. This is because, as described in Section \ref{sec:variable_sources}, our AGN selection method removes systems that have very close neighbours that might affect the flux of the source in question. This may, in turn, artificially lower the AGN interaction fraction. Out of the systems that have been removed from the AGN selection, three involve two nearby systems which exhibit tidal features. While we cannot confidently include these systems in our AGN sample, it is worth noting that, if they were indeed AGN then the interaction fraction in our sample would increase to 0.27 $\pm$ 0.06, with the z-test evaluation suggesting a marginal result which has a p-value of 0.1.

In other words, while the interaction fractions in the AGN and control galaxies are formally consistent within the errors, the issue noted above means that we cannot rule out the possibility that the interaction fraction in the AGN could actually be larger by almost a factor of 1.6 compared to the control sample. Some caution is therefore required when using this relatively small sample size to interpret the interaction fractions. Nevertheless, the relatively low AGN interacting fraction and the fact that this is relatively similar in the AGN and controls, suggests that most of the AGN triggering, at the least in the sample studied here, may be unrelated to galaxy interactions and could be driven by secular processes such as gas accretion and disk instabilities \citep[e.g.][]{Zhao2022,Omori2025}.

In general, variability-based AGN selection can slightly favour early-type hosts, which are dominated by old stars, while a variable nucleus can be harder to detect in bright late-type hosts, where clumpy star-forming regions can introduce spurious stochastic variability which may dilute the nuclear variability \citep[e.g.][]{Baldassare2018,Cleland2022}. Nevertheless, variability-based selection is primarily sensitive to the AGN component, and is therefore generally less biased than other techniques towards the host morphology. Indeed, the similar ETG fractions in the AGN and controls found here suggests that an optical variability-based selection is not biased towards AGN hosts of a particular morphology. 

\begin{figure*}
    \centering
    \includegraphics[width=0.45\linewidth]{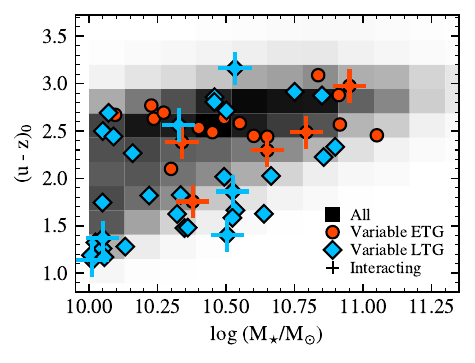}
    \includegraphics[width=0.45\linewidth]{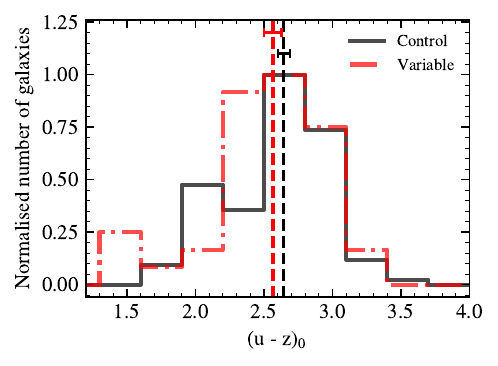}
    \caption{\textbf{Left:} Rest-frame $(u-z)$ colour vs stellar mass. The heatmap indicates the parent COSMOS2020 galaxy population. AGN are shown using filled symbols, with the different morphological classes shown colour-coded (see legend). Interacting galaxies are indicated using crosses. \textbf{Right:} Distributions of the rest-frame $(u-z)$ colours of our AGN and control samples. The dashed vertical lines represent median values while the error bars indicate the uncertainty on the median, calculated via bootstrapping.}
    \label{fig:colour_mass}
\end{figure*}

\begin{figure*}
    \centering
    \includegraphics[width=0.45\linewidth]{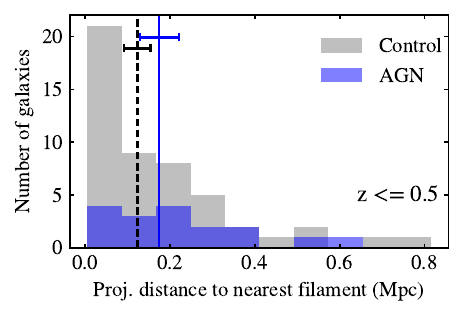}
    \includegraphics[width=0.46\linewidth]{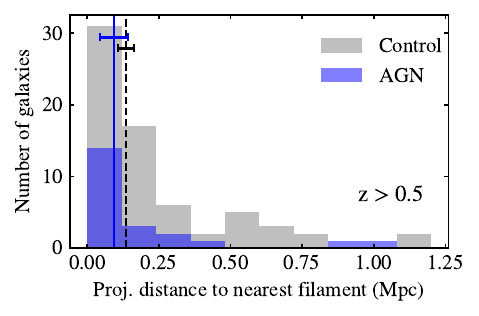}
    \caption{Projected distance to the nearest filament of the AGN population (blue) and the control population (grey). Since the cosmic web evolves with time, we study low ($z<0.5$; left-hand panel) and high ($z>0.5$; right-hand panel) redshift galaxy populations separately. Median values shown using the solid vertical lines and uncertainties in the medians, estimated via bootstrapping, are shown using the horizontal error bars.}
    \label{fig:environment}
\end{figure*}

The similarity in the morphological and interacting fractions found here appears consistent with some studies of X-ray selected AGN when compared to mass-matched control samples \citep[e.g.][]{Fan2014, Kocevski2012, Villforth2014}. Our results are also in agreement with the findings of \citet{Zhong2022}, who use non-parametric methods to study the host morphology of optical-variability-selected AGN at $0 < z < 3$ in the COSMOS field. They find that their sample exhibits nearly equal fractions of early-type and late-type hosts. Similar results have also been reported by \citet{Polimera2018}, who study variability-selected AGN in the mid-infrared and show that the fraction of objects which exhibits strongly disturbed morphologies indicative of mergers is similar to that in a control sample. In the same vein \citet{Bohm2013} also find that the morphology of AGN and quiescent galaxies, traced using morphological parameters, are similar.


\subsection{Rest-frame colour}

We proceed, in Figure \ref{fig:colour_mass}, by studying the star-formation sensitive rest-frame $(u-z)$ colour of our galaxies, which offers a proxy for the recent star formation history. The left-hand panel of Figure \ref{fig:colour_mass} shows rest-frame $(u-z)$ vs stellar mass, while the right-hand panel of this figure shows the distributions of the rest-frame $(u-z)$ colours of our AGN and control samples. 

We find that the rest-frame $(u-z)$ colours of the AGN and controls are similar. The median values overlap within the uncertainties and a KS test between the two distributions yields a p-value of 0.48, indicating that it is unlikely that the two populations come from different parent distributions\footnote{Recall that the null hypothesis, i.e. that the samples are drawn from the same parent distribution, is rejected at a 95 per cent confidence level if the p-value is less than 0.05. Thus p-values greater than 0.05 indicate that the two samples are likely to be drawn from the same parent distribution.}. This suggests that both populations have similar recent star formation histories. This result is similar to that reported by \citet{Xue2010}, \citet{Rosario2013} and \citet{Heinis2016}, who also report a similarity in the rest-frame colour distributions of mass-matched samples of AGN and non-AGN galaxies. Similarly, \citet{Azadi2017} find similar SFR distributions in AGN, selected via X-rays, optical emission line ratios and infrared methods, and inactive galaxies. They find that once selection effects due to stellar mass are accounted for, the SFR distributions of AGN host galaxies are similar to those of inactive galaxies. While infrared and optically-selected AGN can show minor SFR differences due to dust content and selection bias, the overall SFR distribution for mass-matched active and inactive galaxies is comparable. 


\subsection{Local environment} 
\label{sec:cosmic_web_location}

We complete our study by exploring the projected distances to filaments, which we use as our proxy for local environment. Figure \ref{fig:environment} presents the distributions of this quantity for the AGN and control populations. Since the cosmic web evolves with time, i.e. matter moves from less dense to more dense environments and filaments get more defined with time \citep[e.g.][]{Cautun2014}, we split our analysis into two redshift ranges ($z<0.5$ and $z>0.5$). In both redshift ranges the AGN and control distributions are similar, with the median values overlapping within the uncertainties. KS tests between the AGN and controls yield p-values of 0.36 for $z < 0.5$ and 0.65 for $z > 0.5$, suggesting that they are likely to be drawn from the same parent distributions. The results of the KS test are confirmed by a Cram\'er--von Mises test which produces p-values of 0.28 for $z < 0.5$ and 0.91 for $z > 0.5$.  

Our results are consistent with some studies in the recent literature. For example, \citet{Rembold2024} find similarities between the environments of around 300 AGN and non-AGN, where the AGN are selected via a combination of optical emission lines \citep[e.g.][]{Baldwin1981} and the WHAN diagram \citep{Cid2010}. This is similar to the findings of \citet{Cisternas2011} and \citet{Villforth2014}, who find that AGN at low redshift are likely to be triggered through stochastic processes, minor mergers and bar instabilities. In a similar vein, \citet{Wethers2022} have analysed around 200 quasars, spanning low to intermediate redshifts ($0.1<z<3.5$), and find that both the quasar and control populations inhabit similar environments. Finally, \citet{Karhunen2014} report similar galaxy number densities around AGN and luminosity-matched controls at $z<0.5$ within a projected 1 Mpc radius.


\section{Summary}
\label{sec:summary}

We have studied a population of 56 AGN in massive galaxies, identified via their optical broadband variability in the VST-COSMOS survey. VST-COSMOS offers a nearly-identical single visit depth ($r$ $\sim$ 24.6 mag) and temporal baseline (11 years) as the forthcoming LSST, in a 1 deg$^2$ footprint (which is four orders of magnitude smaller than that of the LSST).

The results of this study are therefore a useful precursor to the kind of work that could be performed using LSST in the near future. We have compared the properties of our AGN to a control sample that is matched in redshift and stellar mass in order to study how the behaviour of AGN may vary from the general galaxy population. Our main conclusions are as follows:

 \begin{itemize}
 
    \item The fraction of ETGs in the AGN population (0.55 $\pm$ 0.08) is similar to that in the controls (0.48 $\pm$ 0.04).
    
    \item The fraction of interacting galaxies, identified via visual inspection, in the AGN (0.23 $\pm$ 0.06) and controls (0.16 $\pm$ 0.03) are also similar. Recall, however, that while the interaction fractions in the AGN and control galaxies are formally consistent within the errors, the removal of systems that have a close interacting companion during the AGN selection process, means that we cannot rule out the possibility that the interaction fraction in the AGN could actually be larger (by around a factor of 1.6 compared to the controls). Nevertheless, the relatively low interacting and ETG fractions, and the fact that these are relatively similar in the AGN and controls, suggests that much of the AGN triggering in our sample is unrelated to galaxy interactions.
         
    \item The AGN and controls share similar distributions of rest-frame $(u-z)$ colours, with a median value of $2.57 \pm 0.07$ for the AGN and $2.65 \pm 0.04$ for the control population, suggesting that both populations have similar recent star formation histories. 

   \item On average, the AGN and control populations are located at similar projected distances from cosmic filaments (i.e. large-scale structure). For AGN at $z < 0.5$, the average distance to the closest filament is 0.17 $\pm$ 0.04 Mpc (0.12 $\pm$ 0.03 Mpc  for the controls), while for those at $z>0.5$, the average distance is 0.09 $\pm$ 0.05 Mpc (0.14 $\pm$ 0.03 Mpc). These findings suggest that the local environment has a minimal role, if any, in the triggering of AGN in our sample.

 \end{itemize}

We conclude our study by considering how our results compare to the findings in the past literature. A scenario in which major mergers are not the key drivers of AGN triggering, especially for lower luminosity obscured AGN (which dominate a sample selected via optical broadband variability), aligns well with theoretical models in which stochastic accretion dominates BH growth, \citep[e.g.][]{HopkinsHernquist2006,Martin2018,Alexander2025}. These results are supported by some empirical studies that have probed low to intermediate luminosity AGN within $z<1$ \citep[e.g.][]{Grogin2005,Gabor2009,Bohm2013} but might be in tension with other work that reports elevated merger fractions in AGN compared to control samples, \citep[e.g.][]{Koss2010,Gao2020}. 

The discrepancies are likely to be largely attributable to selection effects involving luminosity dependence and AGN type. Higher merger fractions are reported in literature among AGN samples with high-luminosity \citep[e.g.][]{Yoon2025} and/or obscuration. These samples are mostly identified at infrared or hard X-ray wavelengths \citep[e.g.][]{Treister2012, Ellison2013, Goulding2018, Bonaventura2025} and likely trace the dust-enshrouded stage of the merger process.

\cite{Villforth2023} has compared several studies that explore that connection between AGN and mergers over a range of wavelengths and luminosities. They report an overall lack of correlation between AGN luminosity and merger fraction in the luminosity range $41 \lesssim \log L_X\,(\mathrm{erg\,s^{-1}}) \lesssim 48.0$ (see also earlier work in \citealt{Villforth2014}), which further supports the idea that internal disk instabilities, minor mergers or bars could drive a substantial fraction of BH growth \citep{Hickox2014}. 
\citet{Villforth2023} further attributes the higher merger fractions in AGN found in some parts of the literature to the smaller datasets that have been in these studies. Conversely, they suggest that radio-selected and red AGN do show elevated merger fractions. 

Given that the low-luminosity AGN population peaks at $z<1$, and is therefore more numerous at low redshift than their high-luminosity counterparts, which peak at $z\sim2$ \citep[e.g.][]{Bongiorno2007}, it is desirable to understand the processes responsible for triggering and fueling these AGN to achieve a comprehensive understanding of galaxy evolution. Our results suggest that continuous, secular processes likely regulate the bulk of BH growth at late epochs. Nonetheless, they do not rule out the possibility that mergers do contribute to some AGN triggering at these epochs (or higher redshifts) or in AGN samples that are not studied in this work. 

The occurrence of AGN in both star-forming and quiescent galaxies, as suggested in our colour comparison results, is indicative of accretion being loosely synchronised with prompt host-wide quenching or starburst events at late epochs \citep[e.g.][]{Hickox2014,Schawinski2015}. These results support a scenario in which, at low redshift, SFR and galaxy growth may be weakly coupled or completely decoupled \citep[e.g.][]{Harrison2017} from AGN activity. This is consistent with studies which show that the probability of hosting an AGN depends more on stellar mass than on galaxy colour \citep{Aird2012}, and that instantaneous AGN activity is weakly correlated with the host galaxy's SFR \citep{Rosario2013}. It is interestingly to note, in this context, that indicators of quenching (e.g. red colours) tend to correlate more strongly with BH mass, which is a measure of the cumulative output of the AGN over its lifetime \citep[e.g.][]{Piotrowska2022}.

While the comparison of AGN to control samples has been common in literature \citep[see e.g. the compilation presented in][]{Villforth2023}, our work has offered some novelties and a complementary perspective on past work. For example, we have used deep HST imaging, which has been shown to recover higher merger fractions in optical studies \citep[e.g.][]{Bennert2008} than in ground-based imaging. Additionally, variability selection allows us to identify AGN that may be missed by other optical selection techniques due to host-galaxy dilution of the AGN emission.

A final caveat worth noting here is that the 1 deg$^{2}$ footprint of the VST-COSMOS survey yields a relatively small sample of AGN. The results presented here will therefore benefit from a wider area survey from which a much larger sample of variability-selected AGN can be identified. As noted above, the characteristics of VST-COSMOS mirror those of LSST, which will offer, at completion, an 18,000 deg$^{2}$ footprint with a similar single visit depth and temporal baseline as the data used in this study. Our study provides a preview of the AGN science possible using the LSST, which will enable revolutionary studies of variability-selected AGN as a function of redshift, stellar mass and environment over a significant fraction of cosmic time. Future work will extend this analysis to higher redshifts to test whether AGN triggering mechanisms evolve with cosmic time, to help clarify the mixed results reported for merger fractions at cosmic noon \citep{Kocevski2012, Hewlett2017, Marian2019} and investigate the dependence on AGN luminosity.


\section*{Acknowledgements}

We are grateful to the anonymous referee for many constructive comments that helped us improve the quality of the original manuscript. BB acknowledges a PhD studentship from the Centre for Astrophysics Research at the University of Hertfordshire. G.~M acknowledges support from the UK STFC under grant ST/X000982/1. SK, IL and AEW acknowledge support from the STFC (grant numbers ST/Y001257/1 and ST/X001318/1). SK acknowledges a Senior Research Fellowship from Worcester College Oxford. DD acknowledges PON R\&I 2021, CUP E65F21002880003, Fondi di Ricerca di Ateneo (FRA), linea C, progetto TORNADO, and the financial contribution from PRIN-MIUR 2022 and from the Timedomes grant within the ``INAF 2023 Finanziamento della Ricerca Fondamentale''. 


\section*{Data Availability}

The data underlying this article will be shared on reasonable request to the corresponding author. The density maps were created using the DisPerSE algorithm which is described in \citet{Sousbie2011}


\bibliographystyle{mnras}
\bibliography{references}




\bsp	
\label{lastpage}

\end{document}